\documentclass[iop]{emulateapj}

\slugcomment{Essay based on a talk given at the 2018 NSF Astronomy and Astrophysics Postdoctoral Fellows Symposium}

\shorttitle{Choose Your Own Adventure}
\shortauthors{L. M. Walkowicz}

\begin{document}


\title{Choose Your Own Adventure: \\
    Developing A Values-Oriented Framework for Your Career}


\author{Lucianne M. Walkowicz}
\affil{Baruch S. Blumberg NASA/Library of Congress Chair in Astrobiology, Washington DC 20540}
\affil{Astronomer, The Adler Planetarium,
    Chicago, IL 60605}




\begin{abstract}
We've all heard how academia evaluates our work: papers, papers, and... more papers. While academic publishing is an essential part of disseminating research results, it is only one activity amongst many that make up a career (or for that matter, a life). The following essay is based on an invited talk for the NSF AAPF Symposium, given at the 2018 Winter AAS Meeting. Here, as in the original talk, I offer some practical advice for thinking about one's work within the larger frame of your personal values and goals. While I will draw examples from my own career, I hope to offer readers guidance in articulating what is important to them, aligning their career choices with those values, and establishing metrics that go beyond the h-index.
\end{abstract}

\section{Background}

When I was asked to speak at the NSF AAPF Symposium back in the fall of 2017, the request was very open-ended. It happened to come at an interesting time for me, in that I had just moved to work at the Library of Congress for a year, where I could devote myself mostly full time to studying the ethics of Mars exploration, which I have been thinking about in my non-copious spare time for a while. As such, I had stepped a bit away from the research I was focused on in Chicago, but I was not yet ready to give a full talk on the work I'm doing at the Library. The NSF AAPF Symposium is also somewhat unique, in that it's not often that one has a chance to speak with an audience oriented towards postdocs and organized specifically around a postdoctoral fellowship, so I thought about what my postdoc self would have liked to hear a talk about. What follows is a slightly edited version of the talk I decided to give. 

As corny as it might sound, this essay is about success-- but success that is defined by you, rather than by the external measuring sticks that those of us in academia are used to measuring ourselves with. A better word for success in this context might be ``fulfillment''. ``Fulfillment'' (as I use it here) is a condition that describes the path you are on, rather than a destination you can reach-- and practically speaking, it is only partially in your control, because it is created both by your choices, as well as by circumstances beyond you. 

My argument is that fulfillment stems from aligning your career path with your personal values, and that a great deal of unhappiness comes from a mismatch in the way you are evaluated versus the way you see yourself generating value in the world. Naturally, I can't tell you what your values are, or what choices you should make-- but I hope to provide a way of thinking about choices: a framework for making decisions and prioritizing what you do that you can apply to your own life, if and where applicable. 

First, a short story: when I was in my first postdoc at Berkeley, I went to a  colloquium given by Matthew Bailes of Swinburne University of Technology. He began the talk by telling us that he was really enjoying his visit, that we all seemed really bright, and then he essentially told us that most of us would no longer be astronomers in a few years. He ran through the numbers of how many grads and postdocs were in the department, how many tenure track faculty jobs there are out there, and made a pretty convincing case that we would mostly not be getting the faculty jobs we were ostensibly trying to get. Matthew has a really great, dry sense of humor, so it was a lightly delivered blow, but literally no one wants to hear that, even if you suspect or know it to be true! 

Once he'd spent the first five minutes of colloquium crushing our dreams, Matthew went on to tell us what happened when he found himself with a career not working out the way he'd planned-- and how he eventually came to found the department he is currently in, which he served as director of for over a decade. His talk really stuck with me, because it had always seemed to me that everyone who'd arrived at a stage in their careers where they were giving colloquia at Berkeley had just breezily sailed their way through to there. I'd never heard anyone so much as acknowledge that they had struggled, I'd never heard anyone express vulnerability, and I'd never heard any discussion of what one does when those struggles arise. After the talk, we postdocs debated the relative roles of choice and chance in Matthew's eventual success, and it seemed to me (and still does) to have been a combination of the two.  

Shortly after that colloquium, I applied for and got the TED Fellowship. Many people have heard of TED Talks, but if you don't know, those TED talks come from a big conference, not unlike the AAS in size, but attended by various movers and shakers from the worlds of Technology, Entertainment, and Design (which is what its acronym stands for). However, it is several orders of magnitude more expensive to attend TED than the AAS, so TED started the TED Fellows program to bring early career people who would never be able to afford it to the conference for free. TED was a great experience for me in a variety of ways, but possibly the biggest outcome was that they teamed me up with a career coach, Jen Sellers, who has patiently spent the last 5-6 years walking me through the dark forest of uncertainty that is being a living human being, and most especially, a postdoc. Just as a good research talk distills down some of what one has learned about the universe, I hope to pay Jen's generosity forwards a bit by distilling some of what I've learned by working with her through my postdoc years and into early faculty years. 

 \section{Starting Points}

This essay is considerably more personal than any other professional writing I have done to date. Since the guidance I am providing is based in my own personal experiences, I'd like to acknowledge that this will not be a cure-all or universal solution. I am starting from a place I call ``So Far, So Good'': this is what works for me, but it doesn't mean that my choices are ``correct'', or that they should be your choices, or even that these choices will be good forever. They are intended to be illustrative rather than definitive. Like I said: so far, so good. 

I would also be remiss not to acknowledge that I am speaking from privilege, in a variety of ways: one being that I am a decade out of grad school and several years out of being a postdoc, so I have achieved some form of success: I am still in astronomy, I have a position that pays me to be in astronomy, and people ask me to give talks. I have other obvious privileges as well-- I am white in a mostly white discipline, and though we are also in a predominantly male field, white, cis-sex women have gained more entry to the field of astronomy than our colleagues from other marginalized backgrounds. While I struggle with both mental and physical disability, both are mostly well-controlled through treatment and both are usually not visible, so in sum, I have it easier than a lot of people. 

However, one of the things that has been a challenge for me I think actually has helped me deal with a life in academia, particularly post-grad school, and that's that I grew up with a lot of instability. I come from a family where I was expected to go to college, but I also went to college to get away from an abusive home, and after that I went to grad school not only because I wanted to be an astronomer, but because I had no plan B. I have been without a safety net for most of my life, which for better or worse, one gets used to. 

In reflecting on some of the choices I've made in my life as well as my career, I think there are two events that shaped my framework for decision making more than others: the first of these is that when I was four, my father dropped dead. He went to work in the morning, felt sick at work, had an aneurysm, and the next time I saw him was in a coffin. Though I was really young, I remember that time very well, and I think this forever cemented in my mind that work might be what you are doing on the last day of your life. 

The second major event is that, in my freshman year of college, I went in for a routine health check up and ended up being diagnosed with a pituitary tumor, which is located smack in the middle of your brain. I won't go into details of what that was like, but much like my dad's death altered my path forever, I learned that circumstances can change much faster than you anticipate. It took me much longer to learn the accompanying lesson that this event should have taught me, which is to prioritize and safeguard your health.  

\section{Determining Your Values}

That concludes the death and illness portion of the essay-- on to sunnier topics, like being fulfilled in your career! As I've said, fulfillment is only partly up to you-- so how are we to wrestle with something that isn't completely ours to direct? I began by telling you that it is essential to align your actions and values, so one way to start is by consciously interrogating the values you hold-- which {\bf are} up to you-- and evaluating decisions based on those values. 

As you read the following, keep in mind these two disclaimers: making your decisions in this way may not result in a career in astronomy. Furthermore, aligning your career path with your values may also not be compatible with a career in astronomy, and as a corollary to that: astronomy may ultimately not be the dream of fulfillment you hope for. 

However, I think that engaging in this way may be helpful for the following reason: unless you create them for yourself, there are relatively few formal opportunities to think about your decisions in a larger framework. In academic astronomy, the main time we do this kind of exercise is in planning our path to completing grad school: we choose a research topic, look at what actions we must take to address that topic, and there is a clearly defined marker of success (completing grad school with the terminal degree of your choice). However, once one leaves grad school, the goal posts move. As a postdoc, you know you are supposed to publish-- because that will make you competitive for a second postdoc, or perhaps for a faculty job. If you're in your second postdoc, you're also supposed to publish, for the same reasons. If you pass go and move on to being junior faculty, you are once again supposed to publish, but now you're also supposed to get grants, teach classes, and mentor students. In some ways it's no wonder that publication record becomes a proxy for productivity-- it's a number, something quantifiable that one can quickly determine. The impact of other activities, by comparison, may seem ``squishier'' and more complicated. 

It is also true that some activities in astronomy are more valued in hiring than others-- a very common complaint I hear is that many junior astronomers are interested in outreach and public communication of their science, but that these activities are either not valued, or worse, public-facing activities are actively discouraged because they are considered a ``distraction''. Another common issue is that people are not valued for software development, even if that software is as (or more!) widely used than a hardware instrument like a spectrograph. The valuing of software has made some progress, but only by being shoehorned into something that is publishable and citable. The relative valuing of various activities is an example of an external measuring stick, a measure that is imposed upon us regardless of whether it aligns with our values. You cannot prevent people from measuring you according to their values, but you can make sure what you do aligns with your own. Remember: work may be what you are doing on the last day of your life. 

\section{Practical Strategies for Building Your Framework} 

So with all that context, let's move onto some practical strategies. It is impossible to consciously align your actions with your values if you don't know what your values are. If you are anything like I was as a grad student and postdoc, you are perhaps vaguely aware of having a set of values, and you may intuitively have a sense of whether you are acting within them or not. Very few people, unless prompted, have to specifically articulate what their values are-- but doing so can be a very useful exercise. For those of you with teaching experience, it's a little like the understanding of a scientific topic you gain when you're required to explain it to someone else, where putting words and specificity to something can bring clarity, or highlight areas that need to be clarified. 

\subsection{Write a Mission Statement}

Identifying and articulating your values is a process, but one can use that process to create a helpful beacon: the mission statement. We usually encounter mission statements in the context of organizations-- for example, here's the Adler's: 
\begin{quote}
{\em The mission of the Adler Planetarium is to inspire exploration and understanding of our Universe.}
\end{quote}
And here is that of the Library of Congress: 

\begin{quote}
{\em The Library of Congress's mission is to support Congress in fulfilling its constitutional duties, and to further the progress of knowledge and creativity for the benefit of the American people.}
\end{quote}

Since we are used to hearing about mission statements in the context of institutions, it might sound weird for you to have one, just for you-- but it's really nothing more than taking a close look at what's important to you, and writing it down. There are a lot of guides for writing personal mission statements online, but I find most of them fairly cheesy, so I'll distill the process into prompts that I find helpful. 

\subsubsection{Questions for Reflection}

One place to start is to think of who you admire, and what qualities you admire in them. Maybe they're some famous scientist, maybe they're a family member, friend, or colleague. What do they embody that you admire? These don't have to be perfect people, they don't have to embody who you want to be in every aspect of their being-- the important thing is to think about {\bf why} you admire them.  

What do you see as your purpose? What is important to you? When you do something that feels like an accomplishment, why does it feel that way? In all of this, it is helpful to be as honest and non-judgmental with yourself as possible-- for example, perhaps you like giving public talks because you feel as though you are helping people understand the universe, but perhaps you also like being on stage and performing for a crowd. One of these is driven by an urge to help others, one by the needs of one's own ego, but it's important to understand both if they are both at play. 

Brainstorm your aptitudes: what do you think you are good at? What do you really enjoy doing? What were you doing the last time you felt ``good-tired'' instead of just ``tired-tired''?

Lastly (and here's where we return to death for a moment): what do you want to be known for? Sometimes, this is phrased as your ``legacy'', which definitely implies you have died but never mind that for a moment-- what would you want your colleagues, friends and family to say about you? 

Now the hard part: much like deciding what your thesis topic is going to be, you now have to take the results of this exercise and turn it into a specific articulation of your intent. It helps to be as succinct as possible, but it's okay if your mission is a couple different statements that complement each other. Think 1-3 sentences, not a page. Sometimes phrasing can help: I personally found it too intimidating to write those sentences with ``My mission is...'' so instead, mine start with ``I want to...'' Phrase it however works for you. 

Finally, the really hard part: get someone else to read it. Your mission statement is for you, and you don't have to declare it publicly or put it on the internet or anything like that-- but again, it resembles teaching in that the more clearly you can explain it to someone else, the more clearly you yourself will understand it. So share what you've written with a close friend, or even make a pact with a good colleague to work on this together. Can you explain what you mean to them? Do you understand it yourself? 

One final word on mission statements: they can change, and they can evolve. Always let what you've written guide you without constraining you. Your mission statement should be a lighthouse, not an anchor. 

\subsection{Make Conscious Choices}

Life as a postdoc is a time of opportunity, decisions, and the massive amount of stress that accompanies both. You probably have a lot more freedom than you did as a grad student-- you have more choice in what you work on, who you work with, what conferences you attend, and ultimately, what jobs you apply to. The downside is that choices can be overwhelming, especially when you are bogged down with trying to get a lot of work done in the limited duration of your postdoc.

In my experience, towards the end of grad school and in early postdoc was also a time of some desperation-- we all hear stories about how hard it is to get a job in the field, and so when I was earlier in my career I definitely took the ``spray of buckshot'' approach to job applications: I applied to anything and everything I could remotely be considered qualified for. It stressed me out, and worse, it wasted my time. I can also tell you now, having hired a postdoc myself, that I am not the only person that does this-- part of the reason that astronomy jobs have pools of 300+ applicants is that people are submitting the same application to a lot of jobs they have not closely evaluated. You can't blame anyone for doing that, because seemingly no one is taught how to decide whether a job is suitable for them or not-- but don't do that. The people reading your applications can tell. 

There are good and bad aspects to casting a wide net, just like there are good and bad things about having opportunities. When you are starting out, it can be helpful to take a ``many seeds in the ground'' approach, because sometimes you can't tell which will germinate and eventually grow into something more. However, the flip side of that is that you can get stuck in a cycle of unfocused pursuit, going after any and all things with equal attention. 

\subsubsection{Questions for Evaluating Opportunities}

To help manage your energy, it can be useful to come up with a series of questions you ask yourself when trying to decide what to pursue. These questions will differ from person to person, and even for individuals there will be times when these questions may not be applicable. However, if you don't have a framework to evaluate your decisions, it can be hard to cross-compare competing decisions, and it can be difficult after the fact to articulate {\bf why} you made those decisions. As an example, here are the questions that I came up with: 

\begin{itemize}

\item {\bf How well does this opportunity fit with my mission and values?} This question is the big one-- the things you spend your time on should serve your mission and goals related to it. 

\item {\bf What is new and different about this opportunity?} Does it overlap with or duplicate any existing things I'm doing? What do I hope to gain as a result of taking this opportunity? Be specific: let's say you are invited to a conference. If you are looking for exposure for a new science result, consider who will be in your audience. If you are looking to start new collaborations, look for overlap with people you are hoping to speak with. As a postdoc there can be value in getting yourself out and seen as much as possible, but there's also a limit to that. A related point here is that your hiring will eventually depend on people outside your subfield, so while specialty meetings that are discipline specific are useful for starting new collaborations or identifying future colleagues/employers, meetings that are broader, like ones around a specific telescope or facility, help expose your work to a wider swath of the population. 

\item {\bf How long do I think it will take to fulfill this commitment, or what is the time demand of this opportunity?} Again, be specific: if you join a users' panel or advisory committee and the duration of your commitment to them is three years, that can be made more granular by looking at how many meetings you will have, whether those meetings are a day long or a couple days long, and whether you will have to travel or not. 

Then, at least in my case: {\bf multiply that time estimate by a factor of 2-3}. Know yourself well enough to know when you have regular mistakes you make, like underestimating how long things take, which is something I do all the time. 

\item {\bf What am I willing to give up for this opportunity?} This question is a rough one, but it's one you must ask. As a grad student, then postdoc, then faculty, you will repeatedly have more things on your to do list than you can handle. When you take on a new opportunity or commitment, there will be a cost-- most of the time, most people don't articulate what that is when they take something on, and they end up bearing the cost in some way they didn't anticipate. It could be something simple, like you will have to work an hour later each day, but often these simple things are not what they seem: does that mean an hour less spent with your friends, partner, or family? Does it mean an hour less sleep? These can be difficult things to quantify and hard to look at when you do, but it's important that you are honest and open-eyed. 
\end{itemize}

\subsection{Applying Your Framework}

Once you have identified your values, and created a list of questions that can help you analyze opportunities that come along, you're ready to start applying this framework. Below, I discuss two brief examples of using the evaluation questions in the previous question to specific situations that arose for me (and may arise for the reader).

\subsubsection{What jobs should I apply for?}
As I previously mentioned, my first few rounds of the job slog, I applied for literally everything under the sun. I spent a lot of my own time writing applications for places of dubious fit for me, and perhaps worse I also wasted the time of my letter writers, and of the people on the other end reading the applications. It wasn't until my last time on the job market that I decided to approach my search with a different tactic. My last round of applications was in the second year of my second postdoc, a year {\em before} I absolutely needed a job. When I ran through the list of questions I'd come up with, I saw that what was different about applying then was that I wasn't absolutely required to apply, so I could be a bit more selective, thus limiting the amount of time I'd have to spend (and thus things I'd have to sacrifice) in Questions 2 and 3. I therefore chose my applications based on a single criterion: I only applied to a job if I could, after reading the job ad (and maybe doing a minimal amount of research on the department, if I wasn't already familiar with them) express in a sentence or two why I was a good fit for the job, and why it was a good fit for me. Job applications, after all, are arguments-- you are arguing for your suitability to a particular position. If I felt like I couldn't make that argument right away, I passed on applying for that job. 

\subsubsection{How do I evaluate a faculty offer?} 
That job application cycle was successful for me, in part because of my approach, but also (and perhaps more significantly) I was fortunate that there were a number of opportunities for which I felt I could make a good argument. Again, there are factors out of your control, and I am writing from a position of privilege. During the process of writing down why I thought certain jobs were a good fit, I realized that I thought I would be happy in a fairly wide variety of settings, from teaching-heavy, undergrad-only institutions, to traditional research universities, to something focused on public communication (like the Adler). I interviewed at a sampling of each of these kinds of institutions, and ultimately got two offers, one from the Adler, and one from a very traditional physics and astronomy department at a research university, which I will refer to as Research University. 

I'd like to take a moment to acknowledge here that one factor beyond your control are the norms of astronomical society. When I was in grad school, and I think this is still true now, getting a tenure-track faculty position at a research university was supposed to be the pinnacle of one's existence as an astronomer. Nobody outright told me that, of course, but I had gleaned as much from spending time in astronomy environments. Creating and asking myself these evaluation questions helped me pay attention to my needs and desires, rather than the clamoring voice of societal pressure. 

Both opportunities were new and different, because they were both faculty offers, so that wasn't a great distinguishing criterion. Time-wise, I knew they were both going to absorb all of my time in one way or another, because they essentially represented a phase change in my career. However, they presented fairly different sacrifices: at Research University, I would be expected to do what traditional faculty do: conduct research, mentor graduate students, teach classes, apply for grants, and probably service work for the university in one way or another. After I got the offer from them, I visited Research University, and asked how success (presumably defined in relationship to getting tenure) was defined there. The response (and I'm paraphrasing here but not by a lot) was ``do a lot of research and publish a lot of papers, don't mess up teaching, and when you have tenure [presumably in 6 years] you can do outreach''. That meant to me that I would need to spend a minimum of 6 years {\bf not} doing the public communication that had become very important to me, and I kept thinking, ``what if I get hit by a bus in Year Five?''

The Adler, on the other hand, had its own sacrifices: while the positions are not term limited, they are also not tenure track. It's also a non-profit institution and on a 12 month, rather than 9-month, salary-- which means the job is less stable and the pay is lower than what I would have the potential to earn at a university. In general, your starting salary is one of the biggest influences on your lifetime earnings, so that also means lower lifetime earnings overall. 

So: was I willing to sacrifice the part of me that likes spreading scientific knowledge beyond the experts, or the part that wants to never be in danger of eviction again? In the words of A Tribe Called Quest, ``Riding on the train with no dough, sucks''-- but ultimately, the decision came down to values. I decided that I would rather devote the majority of my time convincing as many people as possible that science is accessible to them if they want it, than to focus on creating new astronomers by shepherding graduate students through research programs. The Adler clearly offered me the opportunity to not only speak with the broader public, but specifically to reach young people of color via the Chicago Public School system field trips, and our outreach programs throughout the city. At Research University, I would largely be reaching kids who had already made it to college, or to grad school. Perhaps most importantly of all, my success at the Adler would be evaluated along similar criteria to what I considered important for myself, demonstrating the alignment of the institution's values with my own. 

\subsection{Evaluation Metrics: Measuring Your Progress Based On Your Values}

As I mentioned at the beginning of this talk, it isn't really all that surprising that publications have become the de facto metric by which success in the field is determined: it's a number, something that can quantify the impact of your research output, something simple enough that it can be calculated by Google Scholar or any number of online tools, and then used to cross-compare between astronomers in very different subfields. As a side note, astronomy has a particularly high publication rate compared to other sciences, which makes it an easy metric too. 

One of the things I have struggled with is that academic publishing is probably my least favorite thing about astronomy-- I think it's important, I like doing research, and I like writing, but I am not great at taking papers to completion (fortunately, I have great collaborators who are better at this than I am). So, I am perhaps a bit biased-- but also, I find it to be a poor proxy for my own output in the field: a H- or i-index doesn't capture things like writing or speaking for non-specialist audiences, creating public science programs, running workshops, managing collaborations, mentoring students, or really most of what I actually spent my time on. Even though I could recognize that this was true, however, I spent many years feeling like a failure, not only as a postdoc but also as faculty, because I didn't think I published ``enough''. On the other hand, I was also enjoying and feeling fulfilled in the things I actually spent time on, but always with guilt and fear hanging over me. 

In the past year or so, it's come to me that while I had a way of making decisions about what to do within that value-oriented framework at the beginning of an activity, I lacked a way of evaluating whether my continuing actions continued to be in line with that framework. A rising H-index, by comparison, is supposed to tell you that you are doing well-- so in short, how do {\bf I} know I'm doing well? 

Coming up with metrics for your progress is difficult, in part because success and fulfillment are qualities of evolving pathways, rather than destinations or fixed outcomes. An evaluation metric therefore {\bf can} be a concrete outcome-- as in, you could aim to publish a certain number of papers, or give a certain number of talks-- but perhaps more importantly, it could be any kind of indicator that is the outward manifestation of your value system. Thinking of traditional metrics of success works within this framework too-- beyond publishing, you could have a certain number of grant applications, or if you're bold, a certain grant acceptance rate could be a metric for you. The idea here though is to expand upon those to be inclusive of non-traditional metrics as well. 

\subsection{Example (Personal) Metrics}

To provide some concrete examples, here is a sampling of the metrics I personally work with, along with what they tell me. 

\begin{itemize}
\item  {\bf Referrals of people in need (usually students)}: One of the things that has started to happen in the past few years is that people have started to come to me for advice. Indeed, that is where this essay comes from. A significant career health indicator for me, then, is how often people not only come to me for help, but how often they refer others to me as well. What this is an indicator of to me is the broader astronomical community valuing my expertise-- it is an acknowledgement that I have a contribution to make. 

\item {\bf Student mentoring}: When I originally took my position at the Adler, I knew that I was sacrificing, in large part, the opportunity to work closely with undergrad and grad students. However, that didn't turn out to be the case completely-- I ended up working with a graduate student from a local institution (Daniel Giles of Illinois Institute of Technology)-- and following that, ended up founding the LSSTC Data Science Fellowship Program for graduate students. I also serve as an unofficial auxiliary mentor via the kinds of referrals I mentioned before, which sometimes evolve into continuing mentoring relationships, and sometimes those referrals are for people facing specific crises. If I am working with students, especially to solve problems, I can be sure that I am providing functional, concrete help to those people-- which is important to me. 

\item {\bf Spaces created for enfranchisement-- of both non-scientists and scientists}: A big motivating factor for me in my work is to help increase accessibility and participation in STEM for diverse peoples, including those who don't necessarily want to be scientists themselves. I therefore think of public engagement-- in any form, whether in the form of a public program through the Adler, or in terms of giving a talk-- as creating a space that empowers people to ask their own questions. Many of us in astronomy focus on disseminating information to people in a way which is accessible, which is a good thing to do, and which is why the primary mode of ``outreach'' is public talks. However, one-way forms of communication don't necessarily leave people feeling empowered to question the world around them. One-way forms of communication also don't give us, as scientists, opportunities to listen and learn what people actually need. Part of my goal is also to better understand how to make astronomy-- and broadly, STEM-- better serve people who have been traditionally marginalized. Instead of asking ``how do we get more diversity in STEM'', we should be asking what we offer, instead of assuming the value is there. I therefore look at how often I am able to create space for someone to feel empowered as a health indicator for my work. 

\item {\bf Audience reach}: Audience reach is an easy example of a seemingly ``squishy'' aspect of communication that actually is easily quantifiable: the number of people you reach with your efforts, whether they be focused on peer-to-peer communication, or focused on reaching a broader audience. Your audience is a quantifiable proxy for the potential reach of your message, whatever that happens to be. 

\end{itemize}

In addition to audience, many of the above seemingly ``squishy'' things can be quantified by asking simple questions: ``how many?'' ``how often?'' ``for who?'' Thinking of each metric in this way can not only help you understand how well you're doing, it can also help you communicate your actions to others who might be more used to measuring your ``output'' by more traditional metrics. On top of that, you can choose from a world full of options by considering impact ahead of time: think about the likely ultimate outcome of an action in terms of the metrics that matter. 

\section{A Few Final Thoughts}

Of the afflictions common to academic life, feeling out of control and/or like your efforts will never be enough surely rank near epidemic proportions. Having said that, it is important to remember that you do have agency, and more importantly, that your value is not equivalent to your productivity. Playing the game of academia by someone else's rules will not necessarily result in a job-- so you may as well play by your own. Also, the normative ``rules'' of the game can change: if you mold yourself solely by the rules that are normative now, you deny the fact that our community values can, should, and are changing, and that change is driven by {\em people like you}.

I will also point out that an essential part of this process is knowing your relationship to risk: beyond values, a lot of what these decisions come down to is understanding the level of risk you consider acceptable in life. In my own experience, very few things are as safe or stable as they seem, but your mileage may vary. After all, it is also possible to fail: you can come up with a mission statement, personal metrics for success, etc, and you still might not meet your goals despite your best efforts. That is OK-- it might suck, but it happens. If that happens, it bears remembering that this framework is flexible, one that can and should be revisited and revised over time. 

One of the powers here is that having a personal framework also helps you identify fit {\bf beyond} the field of astronomy, too-- whether that is in seeking a job industry, or in communicating what you actually do all day to friends and family outside astronomy. While within the field, we often strongly emphasize specialization, the truth is that you are likely more flexible and adaptable in your skills and abilities than your research training would lead you to believe. 

\acknowledgments
Many thanks to the organizers of the NSF AAPF Symposium-- Cameron Hummels, Abby Crites, and Phil Rosenfield-- for inviting me to speak and thus prompting the creation of this work. This essay (and my whole life) has deeply benefitted from conversations with and advice from friends, many of whom read and gave me feedback on the ideas presented here. I would particularly like to thank Jedidah Isler, Ren\'ee Hlo\v{z}ek, Jeff Oishi, Nicole Cabrera Salazar, and Adam Miller, for their friendship, thought-provoking discussion, and insightful comments.






\end{document}